\documentclass[conference]{IEEEtran}
\IEEEoverridecommandlockouts
\usepackage{cite}
\usepackage{amsmath,amssymb,amsfonts}
\usepackage{algorithmic}
\usepackage{graphicx}
\usepackage{textcomp}
\usepackage{xcolor}

\usepackage{amsmath,url}

\usepackage{cite}
\usepackage{amsmath,amssymb,amsfonts}
\usepackage{algorithmic}
\usepackage[]{graphicx}
\graphicspath{{Figures/}}
\usepackage{subfig}
\usepackage{lipsum}
\usepackage{balance}
\usepackage{textcomp}
\usepackage{xcolor}
\usepackage{multirow}
\usepackage{color}
\usepackage{xcolor}
\usepackage{algorithm} 
\usepackage{algorithmic}
\usepackage{listings}
\usepackage{courier}
\usepackage{float}

\def\BibTeX{{\rm B\kern-.05em{\sc i\kern-.025em b}\kern-.08em
    T\kern-.1667em\lower.7ex\hbox{E}\kern-.125emX}}

\begin{document}

\title{Machine-Learning Techniques for \\Detecting Attacks in SDN}

\author{
Mahmoud~Said~Elsayed$^{1}$,
Nhien-An~Le-Khac$^{1}$,
Soumyabrata~Dev$^{1,2,3}$,
and~Anca~Delia~Jurcut$^{1,2}$
\\
$^{1}$~University College Dublin, Dublin, Ireland\\
$^{2}$~Beijing-Dublin International College, Beijing, China\\
$^{3}$ADAPT SFI Research Centre, Dublin, Ireland
\thanks{The  ADAPT  Centre  for  Digital  Content  Technology  is  funded  under  the  SFI Research Centres Programme (Grant 13/RC/2106) and is co-funded under the European Regional Development Fund.}
\thanks{
Send correspondence to A.\ D.\ Jurcut, E-mail: anca.jurcut@ucd.ie.
}
}


\maketitle

\begin{abstract}
With the advent of Software Defined Networks (SDNs), there has been a rapid advancement in the area of cloud computing. It is now scalable, cheaper, and easier to manage. However, SDNs are more prone to security vulnerabilities as compared to legacy systems. Therefore, machine-learning techniques are now deployed in the SDN infrastructure for the detection of malicious traffic. In this paper, we provide a systematic benchmarking analysis of the existing machine-learning techniques for the detection of malicious traffic in SDNs. We identify the limitations in these classical machine-learning based methods, and lay the foundation for a more robust framework. Our experiments are performed on a publicly available dataset of Intrusion Detection Systems (IDSs).
\end{abstract}

\begin{IEEEkeywords}
Software Defined Networking, Intrusion Detection Systems, Denial of Service (DoS), machine learning.
\end{IEEEkeywords}

\section{Introduction}
Software Defined Network (SDN) is extensively used in different network organizations in the last few years. It can facilitate the network programmability through decoupling the control plane from the data plane. This separation simplifies the ability to deploy new resources quickly with less hassle. 
Hence, the programmability feature of SDN inspires many commercial companies to use SDN in their cloud network. It allows the organizations to achieve the required behavior by implementing new applications programming interface (API)  to meet their needs. SDN architecture is comprised by three distinct layers: data plane, SDN controller, and application layer. 



Despite the great success of SDN in several organizations, the separation of the control plane from the underlying data plane increases the chance of exposing the SDN platform to many threats and risks~\cite{yan2015software}. Therefore, the SDNs are vulnerable to different network attacks, including volumetric attacks \emph{viz.} SYN floods, service-specific attacks and Denial of Service (DoS) attacks. Such attacks can overwhelm the different  three  layers  of  SDN  with security concerns. Many defense methods are conducted to detect and mitigate the DoS attacks in the SDN network, such as presented in ~\cite{yuan2016defending}. 
These techniques are based on the collection of the flow features from received packets like the average number of packet, the frequency of IP addresses, average number bytes, \emph{etc.} as significant features to detect the DoS inside the SDN network. In the case of malicious traffic, the size of monitored packets will be higher than the previous defined threshold or decision boundary.
Although the existing techniques do not require high processing resources to manage and control the data flow; it cannot protect the network for a long time. The rapid growth in the generated data and the huge number of packets that need to be analyzed and classified in the real-time are the main drawbacks in the current scenario. 
Owing to the similarities between benign traffic and malicious packets, it becomes a challenging task to detect the malicious traffic. The attacker can easily modify the packet header to look like normal traffic and deceive the detection systems without being discovered. 
Since anomalies are continuously evolving, simple modifications can be made that will allow the attacker to exploit this technique with new kinds of attacks.

Recently, machine learning (ML) and data mining techniques are playing an essential role in the detection and the classification of intrusion attacks. Several machine learning studies have been conducted in different domains~\cite{dev2016ground,nwosu2019predicting}; however only a few of these are on SDN. 
There are many issues that can influence machine learning performance, such as the feature selection methods, the dataset used \emph{etc.} However, there are no related works that perform a systematic analysis of these machine-learning techniques. The main contributions of this paper are as follows: (a) we perform a methodical analysis of the popular approaches for detecting attacks in SDN, and provide benchmarking analysis of traditional machine-learning based approaches; and (b) we provide a feature analysis of the input feature space of the dataset, and provide recommendations for a reliable intrusion detection system.


\begin{table*}[htb]
\centering
\footnotesize   
\begin{tabular}{p{2.5cm}|p{5cm}|p{2.5cm}|p{2.5cm}|p{2.5cm}|p{0.5cm}}
\textbf{ML algorithm} & \textbf{Features used} & \textbf{Dataset used} & \textbf{Dataset size} & \textbf{Classes} & \textbf{Reference} \\
\hline 
SVM, Naive Bayes, KNN & number of Packets, Protocol, Delay, Bandwidth, Source IP and Destination IP
 & Simulation based & 6000 data instances, out of which 2000 are genuine and 4000 are malicious packets & Normal and anomaly & \cite{prakash2018intelligent} \\
 
SVM & srcIP, srcPort, desIP, desPort, proType & Simulation based & -- & Normal and anomaly & \cite{li2018using}\\

SVM & speed of source IP (SSIP), speed of source port (SSP), Standard Deviation of Flow Packets (SDFP), Deviation of Flow Bytes (SDFB), speed of flow entries (SFE), Ratio of Pair-Flow (RPF) & Simulation based & -- & Normal and anomaly  & \cite{ye2018ddos}\\

ASVM & Average number of Flow  packets, Average number of flow bytes, Variation of flow packets in the sampling interval, Variation of flow bytes  per in the sampling interval, Average duration of traffics & Simulation based & 300 Samples  & Normal and anomaly & \cite{myint2019advanced}\\


Hidden Markov Model (HMM) & length of the packet,  source port,  destination port, source IP address, and destination IP address & Simulation based & -- & Normal and anomaly  & \cite{hurley2016hmm}\\

K-means , SVM  & packet count, byte count, duration & Simulation based & 4400 & Normal and anomaly & \cite{da2016atlantic}\\

Random Forest  & ten feature sets & KDD99 & 494,020 pieces,10\% training set & Normal and anomaly & \cite{song2017machine}\\

C4.5, Bayesian Network (BayesNet), Decision Table (DT), and Naive-Bayes & -- & LongTail & 278,598 & Normal and anomaly & \cite{nanda2016predicting}\\

SVM & Used 23 features, 29 features and another experiment apply all 41features & KDD99 & -- & Normal and anomaly & \cite{wang2016efficient}\\

Naive Bayes, K-Nearest neighbour, K-means and K-medoids  & Only one feature & Taken from traced file obtained from TCP traffic between Lawrence Berkley Laboratory and
the rest of the world.
 & -- & Normal and abnormal & \cite{barki2016detection}\\

SVM & 25 from 41 features & NSL-KDD & 125973 & Normal and abnormal & \cite{alshamrani2017defense}\\

Hard thresholds With fuzzy control
 & Distribution of Inter-Arrival Time, Distribution of packet quantity per flows, Flow quantity to a server, Number of source IP addresses to a server, Total traffic volume counted in bytes to a server.
 & Log files collected from NetNam (Vietnamese ISP) & -- & Normal and abnormal & \cite{smith2017securing}\\
\hline 
\end{tabular}
\caption{Summary of related works for both ML-based simulation and ML-based public datasets.}
\label{table:summary}
\end{table*}

\section{Detecting Attacks in SDN}

There are two main approaches based on machine learning that are currently used to detect exploitable attacks on SDN network: (a) approaches based on simulation, and (b) approaches based on public datasets. In this section, we provide a classification and a discussion of the existing methods for both of these approaches. Table~\ref{table:summary} summarises the related works.

\subsection{ML based simulation}

\begin{figure}[htb]
  \begin{center}
    \includegraphics[width=0.27\textwidth]{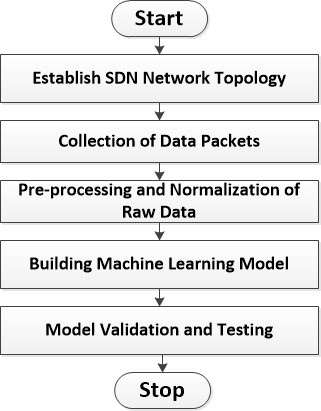}
  \end{center}
  \caption{The graph shows the sequential steps to create simulation dataset in SDN Network.}
  \label{fig:Simulation-plot}
\end{figure}

The work presented in \cite{prakash2018intelligent,li2018using,ye2018ddos, myint2019advanced, hurley2016hmm, da2016atlantic} is using network simulation-based techniques to 
detect the malicious traffic in SDN network. Fig.~\ref{fig:Simulation-plot} discusses the sequential steps in generating a simulation dataset. In these approaches, researchers established network topology with legitimate hosts to generate normal traffics, and other hosts act as pots to create attack traffics. They used public tools like \texttt{Scapy} or \texttt{hping3}, to simulate DoS attacks. The characteristic features, such as speed of source IP or speed of source port, flow packets \emph{etc.} are extracted from the collected traffic for normal and malicious data separately. Then, all of these samples are random shuffling in a \texttt{.CSV} file to create the row data which are used in the training model. The learner model can be used later to classify the normal and intruded packets inside the SDN platform. 

It is clear that these approaches are fast in its computation, and simple to analyze. 
However, it has many restrictions that can be summarized in the following discussion.

Firstly, the created dataset has a very small size and therefore, it is not enough to give accurate results (number of samples are few hundreds or few thousands in most cases). Additionally, the generated data contains only a few kinds of attacks. These attacks are not realistic to represent the diversity of anomalies that are present on the internet. The limitation of diversity in attack types  helps the attacker to know the normal behavior of the detection mechanism, and it can easily craft an attack to replicate this behavior. 

Secondly, the number of extracted features are insignificant, and the small number of features are not enough to cover the behavior of all attacks. Furthermore, the used features for the learner module were extracted primarily from the packet header without further inspection of the payload data. The header field can be easily modified to look like normal ones. Typically, the attacker can easily implement malicious code inside the payload packets to cheat the detection systems such as Root to Local (R2L) attacks, and worm infections. As a result, they return a poor accuracy for detecting application-level attacks.

\subsection{ML based public datasets}

The work introduced in \cite{nanda2016predicting,smith2017securing, barki2016detection,song2017machine, wang2016efficient,alshamrani2017defense} is using 
public dataset to detect intrusion inside SDN environment. The selection of the proper dataset has a great impact in the evaluation of network intrusion detection systems (IDS).  Unfortunately, most of the publicly available datasets are not realistic, and they lack variety in the type of attack to cover all trends found in the internet today. One of the main reasons for these shortage returns to privacy and legal issues for the service providers to publish their network data. As a result, most of available datasets fail to give acceptable accuracy when deployed with intrusion systems.  There are several datasets  such as KDDCUP'99~\cite{KDD99}, CICIDS2017~\cite {sharafaldin2018toward }, ISCX2012~\cite{ ISCX2012}, Kyoto~\cite{song2011statistical}, LBNL~\cite{ nechaev2004lawrence}, UMASS~\cite{ nehinbe2011critical}, CDX~\cite{ sangster2009toward},  ADFA~\cite{ creech2013generation}, DEFCON~\cite{ DEFCON00} have been used  for intrusion detection systems. For simplicity, we will evaluate different machine learning algorithms on NSL-KDD dataset, which is the modified version of KDDCUP'99 and produced to overcome the inherent problems of the KDDCUP'99 data set such as redundancy problem.

\section{Results and Discussions}
In this section, we present our experimental results on a dataset of network attacks. We illustrate the non-linearity of the input feature space, and also benchmark several machine-learning based approaches on this dataset.

\subsection{Dataset}
The KDDCUP'99 is the original dataset that was mainly used by researchers for benchmarking purposes~\cite{shone2018deep, dawoud2018deep}. This dataset was generated from the tcpdumps of DARPA Intrusion Detection System (IDS), and was created in Lincoln Lab. However, there are inherent several drawbacks in this dataset. Most of the observations in KDDCUP'99 dataset are highly redundant, and therefore similar observations occur in both training and testing set. In order to mitigate the drawback of this dataset, a new variant called NSL-KDD dataset~\cite{tavallaee2009detailed} was released by Tavallaee \emph{et al.}. This new dataset do not contain the inherent demerits of KDDCUP'99, and is now used as the \emph{de facto} benchmarking dataset by all researchers. 

The NSL-KDD dataset contains $24$ different type of attacks in its observation records. The dataset is divided into training set and testing set. They are referred as KDDTrain+ and KDDTest+ for training and testing set respectively. A smaller subset comprising $20\%$ of the entire records are present as separate files. They are called KDDTrain-- and KDDTest-- for training and testing sets respectively. In this paper, we consider all the attacks in a combined category of malicious traffic. We formulate our objective in this paper as a binary classification problem, wherein each observation in the NSL-KDD dataset is categorized as \emph{legitimate} or \emph{malicious} traffic.

\subsection{Non-linearity of the feature space}

Prior to the application of any machine-learning techniques, it is important to analyse the high-dimensional feature space generated by the dataset. We generate the original feature space by converting all the features in the dataset into corresponding feature vectors. We normalize all numerical features of the dataset in the range $[0,1]$. The categorical features are converted to vectors using one-hot encoding.  

\begin{figure}[htb]
  \begin{center}
    \includegraphics[width=0.45\textwidth]{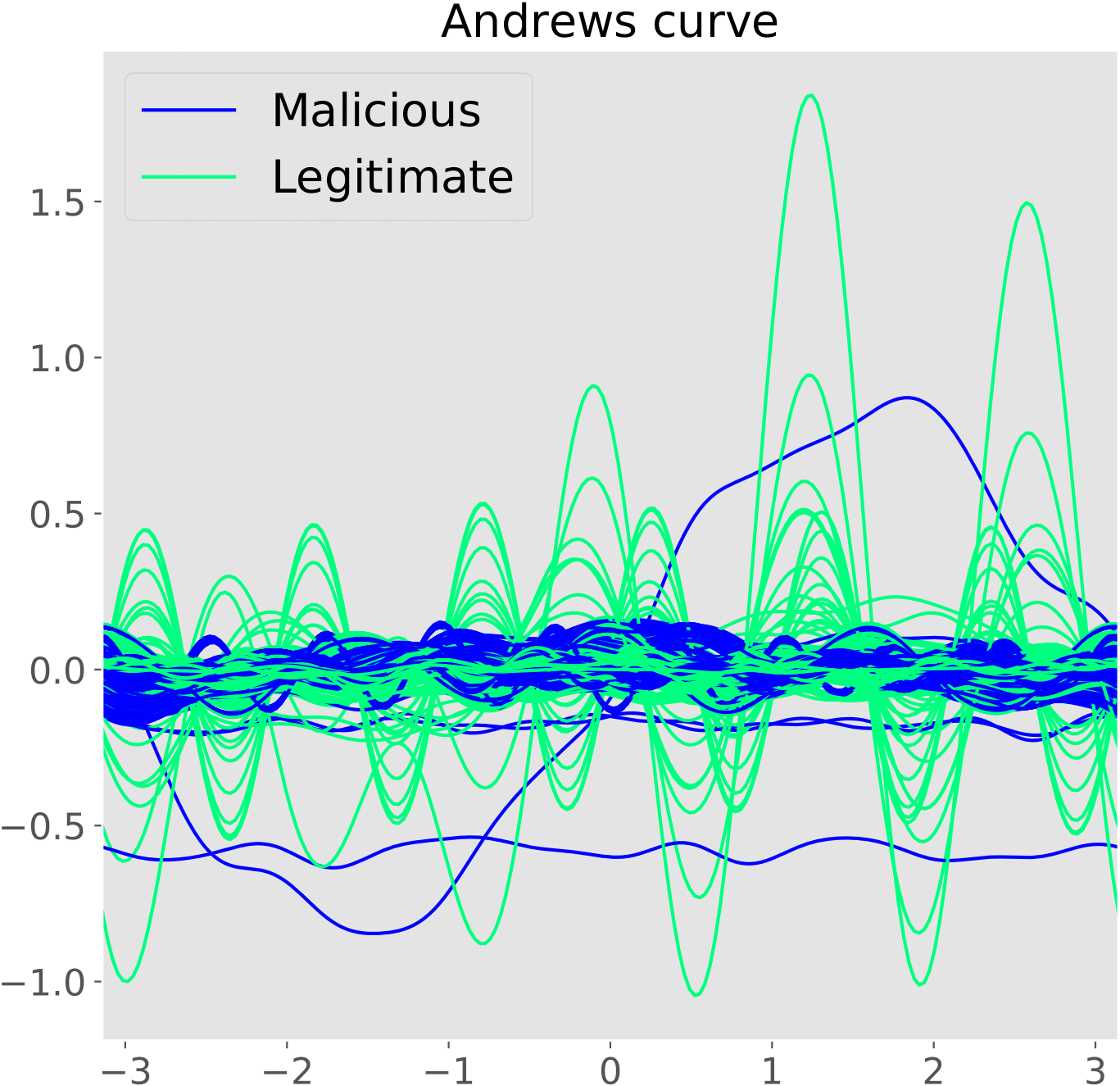}
  \end{center}
  \caption{We plot the Andrews curve for the NSL-KDD dataset. We observe that the data streams are highly intertwined with each other.}
  \label{fig:and-curve}
\end{figure}

We attempt to analyse the non-linearity in this generated feature space by computing the Andrews curve in the NSL-KDD dataset. The Andrews plot is essentially a visual representation of the original high-dimensional feature space into two dimensions. It is computed by representing the original feature space as a finite Fourier series. Fig.~\ref{fig:and-curve} describes the Andrews curve plotted for the NSL-KDD dataset. Each observations in the dataset is represented by the individual curve. These individual curves are color coded based on the type of the traffic -- \emph{malicious} and \emph{legitimate}. We clearly observe that the two type of curves are not clearly separated. The individual streams from the two categories are intertwined with each other. This indicates that there exists a high-degree of non-linearity in the feature space.

\begin{figure}[htb]
  \centering
	\includegraphics[width=0.48\textwidth]{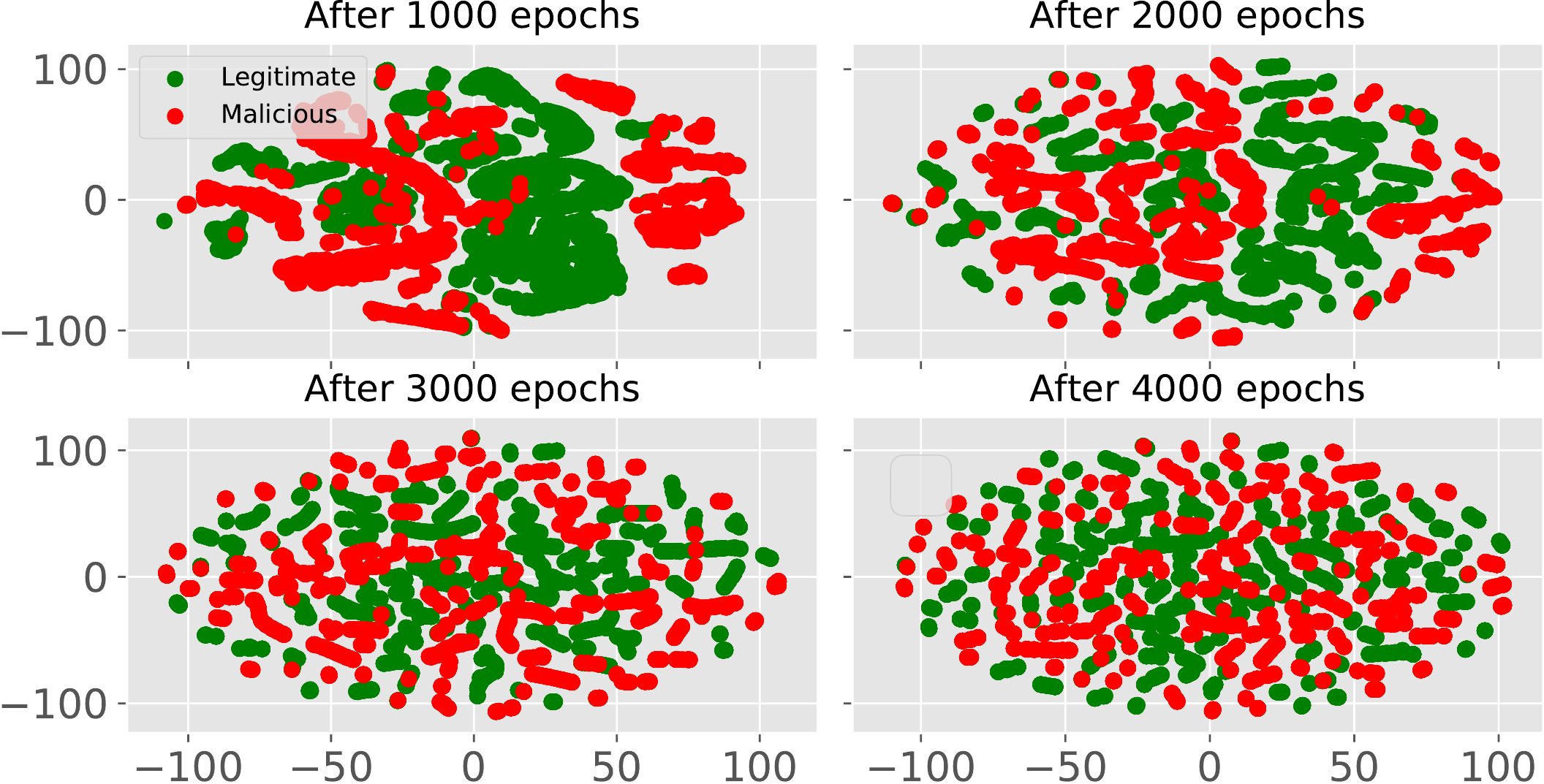}
  \caption{We show the separation of normal and malicious traffic using a t-SNE plot. As t-SNE is a non-linear dimensionality reduction technique, we observe that the separation of the two classes is better in the two dimensional subspace.}
	\label{fig:tsne-plot}
\end{figure}

Furthermore, in order to further investigate the non-linearity nature of the feature space, we generate the t-Distributed Stochastic Neighbor Embedding (t-SNE) plot. Fig.~\ref{fig:tsne-plot} illustrates this. The t-SNE plot is a popular technique of dimensionality reduction technique, that is mostly used for visualizing high-dimensional feature space. We use this technique as a non-deterministic measure to cluster the malicious and legitimate traffic. We observe from Fig.~\ref{fig:tsne-plot} that the observations belonging to two classes are fairly separated after $1000$ epochs. The data observations have a less overlapping nature for higher number of epochs, and are clustered in well-defined regions. This clearly indicates the presence of non-linearity in the original feature space. We can separate the two classes in the dataset, \emph{only} via the use of non-linear dimensionality reduction techniques \emph{viz.} t-SNE. t-SNE takes a long time for its computation to separate the different samples.  We use Principal Component Analysis (PCA) to reduce the original feature dimension space from $122$ to $20$ features, to avoid any noise and speed up the t-SNE calculations. 

\begin{table*}[htb]
\centering
\normalsize 
\begin{tabular}{r||l|l|l|l|l}
\textbf{Dataset} & \textbf{Approach} & \textbf{Precision} & \textbf{Recall} & \textbf{F-score} & \textbf{Accuracy [in \%]} \\
\hline
\multirow{5}{*}{KDDTrain+ and KDDTest+} & SVM~\cite{tian2012recent} & 0.80 & 0.75 & 0.75 & 75.3 \\
 & J48~\cite{kaur2014improved} & 0.85 & 0.81 & 0.81 & 81.5 \\
 & Naive Bayes~\cite{rish2001empirical} & 0.80 & 0.76 & 0.75 & 76.1 \\
 & Random Forest~\cite{goel2017random} & 0.85 & 0.80 & 0.80 & 80.4 \\
 \hline 
\multirow{5}{*}{KDDTrain-- and KDDTest--} & SVM~\cite{tian2012recent} & 0.75 & 0.53  & 0.61 & 52.7 \\
 & J48~\cite{kaur2014improved} & 0.83 & 0.64 & 0.68 & 63.9 \\
 & Naive Bayes~\cite{rish2001empirical} & 0.76 & 0.56 & 0.61 & 55.7 \\
 & Random Forest~\cite{goel2017random} & 0.84 & 0.63 & 0.67 & 62.9 \\
\hline 
\end{tabular}
\caption{Evaluation of benchmarking machine-learning methods in NSL-KDD dataset. We evaluate their performance by computing the precision, recall, F-score and accuracy values.}
\label{table:benchmarking}
\end{table*}

\subsection{Evaluation}
In this paper, we deal with a binary classification problem, wherein we are interested to detect the malicious traffic in the dataset. Let us suppose that $TP$, $TN$, $FP$ and $FN$ denote the true positive, true negative, false positive and false negative respectively, while detecting attack samples in the dataset. We report the following metrics: precision, recall, F-score and accuracy values for the various benchmarking methods. These metrics are defined as:

\begin{equation}
	\mbox{Precision}=\frac{TP}{TP+FP},
    \label{eq1}
\end{equation}   

\begin{equation}
	\mbox{Recall}=\frac{TP}{TP+FN},
    \label{eq2}
\end{equation}   

\begin{equation}
	\mbox{F-score}=\frac{2\times\mbox{Precision}\times\mbox{Recall}}{\mbox{Precision}+\mbox{Recall}},
    \label{eq3}
\end{equation}   

\begin{equation}
	\mbox{Accuracy}=\frac{TP+TN}{TP+TN+FP+FN}.
    \label{eq4}
\end{equation} 

We benchmark the popular machine-learning methods and evaluate their performance in detecting attacks. We benchmark SVM~\cite{tian2012recent}, J48~\cite{kaur2014improved}, Naive Bayes~\cite{rish2001empirical} and Random Forest~\cite{goel2017random} in the NSL-KDD dataset. Table~\ref{table:benchmarking} summarises the results of the different approaches. We observe that the results on KDDTrain+ and KDDTest+ are in general better than that of KDDTrain-- and KDDTest--. This is because models trained on a smaller dataset fail to converge in the limited number of samples. Also, we observe that J48 is the best performing approach, as compared to all benchmarking methods. 


\subsection{Discussion}
The existing classical ML-based techniques have difficulties in detecting 
sophisticated attacks in large-scale network environments. Several challenges can cause poor classification results because of irrelevant features, and lack of labeled training data. 
These factors lead to the difficulties in detecting the attacks by using the classical ML methods. 

Recently, deep learning (DL) is one of the most significant solution for solving the weaknesses of machine learning~\cite{chalapathy2019deep}. 
At present, many companies, such as Google, Microsoft, Facebook, have broadly used DL in different applications \emph{i.e.} speech recognition, image processing. 
The vital part of DL techniques extract the intensive features automatically from unlabeled data records. Therefore, it can be applied in many cybersecurity tasks such as intrusion detection, traffic analysis \emph{etc.} 
Deep learning can address large-scale network traffic using multiple processing layers to reconstruct the unknown structure in the input distribution. It finds a good representations of the input data, which is generally hard to obtain using traditional methods. Nowadays, deep learning algorithms such as recurrent neural network, deep belief network, restricted Boltzmann machine, and deep autoencoders attract the attention of research community and become a popular topic of research for dimensionality reduction issues. DL models have used in the development of IDS as they are capable of representing high‐level features into more abstract data features. Different studies attempted to select appropriate features for intrusion detection, and subsequently develop a learning model via an algorithm.

\section{Conclusion and Future Work}
In this paper, we have provided a detailed study of various approaches based on classical ML techniques that are used in detecting attacks in SDN. We perform our benchmarking experiments on NSL-KDD dataset, and explained why traditional machine-learning based methods fail to have a good performance. In the future, we will propose a DL-based framework that will have superior performance as compared to the state-of-the-art methods. 


\balance

\bibliographystyle{IEEEbib}

\end{document}